\documentclass[10pt,letterpaper]{article}
\usepackage[top=0.85in,left=2.75in,footskip=0.75in]{geometry}
\usepackage{rotating}
\usepackage{multirow}

\usepackage{amsmath,amssymb}

\usepackage{changepage}

\usepackage{textcomp,marvosym}

\usepackage{cite}

\usepackage{nameref,hyperref}

\usepackage[right]{lineno}

\usepackage[nopatch=eqnum]{microtype}
\DisableLigatures[f]{encoding = *, family = * }

\usepackage[table]{xcolor}

\usepackage{array}

\newcolumntype{+}{!{\vrule width 2pt}}

\newlength\savedwidth

\newcommand\thickhline{\noalign{\global\savedwidth\arrayrulewidth\global\arrayrulewidth 2pt}%
\hline
\noalign{\global\arrayrulewidth\savedwidth}}

\raggedright
\usepackage[aboveskip=1pt,labelfont=bf,labelsep=period,justification=raggedright,singlelinecheck=off]{caption}

\makeatletter
\renewcommand{\@biblabel}[1]{\quad#1.}
\makeatother

\usepackage{lastpage,fancyhdr,graphicx}
\usepackage{epstopdf}
\fancyheadoffset[L]{2.25in}
\fancyfootoffset[L]{2.25in}
\begin{document}
\vspace*{0.2in}

\begin{flushleft}
{\Large
\textbf\newline{Bayesian probabilistic projections of proportions with limited data: An application to subnational contraceptive method supply shares.} 
}
\newline
\\
Hannah Comiskey\textsuperscript{1*},
Niamh Cahill \textsuperscript{2},
Leontine Alkema \textsuperscript{3},
David Fraizer \textsuperscript{1},
Worapree Maneesoonthorn\textsuperscript{1}.
\\
\bigskip
\textbf{1} Department of Econometrics and Business Statistics, Monash University, Melbourne, Australia.
\\
\textbf{2} Department of Mathematics and Statistics, Maynooth University, Kildare, Ireland.
\\
\textbf{3} Department of Biostatistics and Epidemiology, University of Massachusetts Amherst,Amherst, MA, USA.
\\
\bigskip

* hannah.comiskey@monash.edu

\end{flushleft}
\section*{Abstract}
Engaging the private sector in contraceptive method supply is critical for creating equitable, sustainable, and accessible healthcare systems. To achieve this, it is essential to understand where women obtain their modern contraceptives. While national-level estimates provide valuable insights into overall trends in contraceptive supply, they often obscure variation within and across subnational regions. Addressing localized needs has become increasingly important as countries adopt decentralized models for family planning services. Decentralization has also underscored the need for reliable subnational estimates of key family planning indicators. The absence of regularly collected subnational data has hindered effective monitoring and decision-making. To bridge this gap, we propose a novel approach that leverages latent attributes in Demographic and Health Survey (DHS) data to produce Bayesian probabilistic projections of contraceptive method supply shares (the proportions of modern contraceptive methods supplied by public and private sectors) with limited data. Our modeling framework is built on Bayesian hierarchical models. Using penalized splines to track public and private supply shares over time, we leverage the spatial nature of the data and incorporate a correlation structure between recent supply share observations at national and subnational levels. This framework contributes to the domain of subnational estimation of proportions in data-sparse settings, outperforming comparable and previous approaches. As decentralization continues to reshape family planning services, producing reliable subnational estimates of key indicators is increasingly vital for researchers and policymakers. 

\section*{Introduction}

FP2030 is a global partnership committed to empowering women and girls through investments in rights-based family planning \cite{FP2030}. This approach emphasizes accessibility and availability of contraceptives \cite{FP2030rights}. Traditionally, the public sector has led family planning efforts \cite{DeLong2020}, but declining donor funding has spurred interest in private sector engagement \cite{Cahill2018, Cutherell2024, Ross2015}. Leveraging private sector contributions enhances equitable, accessible, and sustainable reproductive healthcare \cite{Chakraborty2018, Weinberger2017}. Recent initiatives, such as WHO’s 2024 online training for pharmacists, aim to expand contraceptive access in low- and middle-income countries \cite{PharmacyWHO2024}. Pharmacies play a crucial role, particularly in providing short-term methods like oral contraceptives and condoms \cite{Corroon2016}. Unmarried women and teenagers disproportionately rely on the private sector due to privacy concerns \cite{Bradley2022}. Therefore, understanding contraceptive access patterns is essential for equity \cite{Weinberger2017, Bradley2022}. National estimates of family planning indicators aid global reporting but overlook local trends and fail to assess if target levels for indicators are being ubiquitously attained across the nation. Furthermore, a decentralized approach to family planning management improves sector performance \cite{Bossert2002, DHSSpatial2021}. Subnational estimates are widely used in demography for tracking contraceptive prevalence, unmet need, and under-5 mortality \cite{Li2019, Mercer2019, Wakefield2019}, yet gaps remain in monitoring contraceptive supply sources. \newline
\newline
Small area estimation (SAE) involves estimating characteristics of sub-populations, where there may be few or no samples available, by 'borrowing strength' from related populations to improve accuracy \cite{Ghosh1994}. Using SAE approaches to model demographic and health outcomes is a well-established practice. In 1979, the foundational Fay-Herriot modelling approach described a two-stage hierarchical area-level model that allowed for the weighted estimates of area-level characteristics \cite{Fay1979}. Building on this, in 1991 Basag, York, and Mollie (BYM) described an approach that uses both spatially structured variation and independent unstructured variation to capture underlying disease prevalence \cite{Besag1991}. Since these models, the area of SAE has exploded with progress in many different directions including generalised linear mixed models, Bayesian hierarchical models, and conditional autoregressive (CAR) models \cite{Chen2014} \cite{Wah2020} \cite{Tessema2023} \cite{Peterson2024}. \newline
\newline
Previous works considered the distribution of contraceptive method supply shares at the national level using a Bayesian hierarchical penalized spline model, and explored its application to subnational data. \cite{Comiskey2024} \cite{Comiskey2023}. Bayesian hierarchical models are also very useful in SAE. They allow for the pooling of information across larger populations to inform smaller sub-population estimates, where data is more sparse. Therefore, Bayesian hierarchical models will be used again in this instance to inform and estimate subnational method supply shares. The Comiskey et al. (2024) approach also estimates and incorporates cross-method correlations to inform the rates of change in spline coefficients. However, at the subnational level, capturing cross-method correlations within the rates of change between spline coefficients becomes difficult due to the heterogeneity of subnational data. Instead, we present an approach that captures the cross-method correlations for the most recently observed survey levels of public sector supply shares. Lastly, the Comiskey et al. (2024) approach uses a three-tiered hierarchical grouping structure, reflecting the geographical nature of the method supply share data, when estimating key model parameters. However, we found that a simplified two-tiered hierarchical estimation process performed just as well while also reducing model complexity.\newline
\newline
Overall, we have adapted the national-level modelling approach of Comiskey et al. (2024) to subnational regions where the sampling density is lower (i.e., fewer women sampled per estimate) and consequently the uncertainty associated with the survey observations is higher than what is observed as the national level. While this updated approach also uses Bayesian hierarchical estimation techniques for key parameters, we employ simplified geographic grouping structure. In addition, the updated approach provides key insights into the latent cross-method correlations that exist between the most recently observed levels of contraceptive method supply shares. Finally, the flexibility of this updated modelling approach captures the nature of an otherwise complex time- and spatially-varying dataset. At the time of writing, these subnational estimates, complete with uncertainty measures, represent a significant advancement in monitoring contraceptive supply shares. They provide an insight into subnational spatial and temporal supply-share trends for researchers, with the potential to improve access and equity in family planning services. \newline
\newline
The outline of this paper is as follows. We begin with the definitions used throughout this paper and the data utilised in this approach. In the Methods section we describe the hierarchical penalised spline model. In Results, we consider the output of this model, which we then compare with various approaches listed in the Supplementary Materials, via an extensive model comparision study. We conclude the paper with a discussion. The Supplementary Materials contain more technical details and additional supporting information.

\section*{Definitions and data sources}

\subsection*{Definitions}
Modern contraceptive methods are defined as ``a product or medical procedure that interferes with reproduction from acts of sexual intercourse" \cite{Hubacher2015}. During this investigation, we consider five main modern methods of contraception. These are female sterilization, oral contraceptive pills (OC pills), implants (including Implanon, Jadelle and Sino-implant), intra-uterine devices (IUD, including Copper- T 380-A IUD and LNGIUS), and injectables (including Depo Provera (DMPA), Noristerat (NET-En), Lunelle, Sayana Press and other injectables). Contraceptive method supply shares are defined as "the percent distribution of the types of service-delivery points cited by users as the source of their current contraceptive method (if more than one source, then the most recent one)" \cite{D41_definition}. This paper considers a public/private sector breakdown for each of the five contraceptive methods listed. Contraceptives supplied by government-funded health facilities and home/community deliveries are considered to be supplied by the public sector, while supplies that come from sources outside the public sector, including enterprises ran for-profit or non-profit, can be defined as coming from the private sector. 

\subsection*{Data sources}
Following the data curation approach described in Comiskey et al., we use a database of administration-1 level Demographic and Health Survey (DHS) observations for public and private sector modern contraceptive method supply shares with their associated standard errors \cite{Comiskey2024}. This database was created on 01/09/2024 using the Integrated Public Use Microdata Series (IPUMS)-DHS data \cite{IPUMS}. The authors did not have access to information that could identify individual participants during or after data collection. The variables contained within the IPUMS-DHS database are consistent over time and space, making them suitable for temporal and spatial analysis. The countries included in this study are selected from those participating in the FP2030 initiative. There are 24 countries containing 192 subnational regions, included in this study. Table \ref{table1} gives a summary of each country included in the dataset. In this dataset, Burkina Faso has the oldest 'latest' survey, collected in 2010. In contrast, Madagascar has the newest 'latest' survey collected in 2021. Recorded in this dataset, 56\% of countries collected their latest surveys between 2015 and 2018, while a further 28\% of countries collected their latest survey in 2014 or before. Afghanistan has one survey available and the largest number of subnational areas, with 29 distinct regions. Malawi, Guinea and Cameroon are divided into 3 subnational areas, making them the countries with the least number of subnational regions. Finally, the amount of survey data available varies per country. Senegal has the largest number of surveys in the dataset with 9 surveys collected between 1990 and 2021, while Myanmar and Afghanistan have only one survey present in the dataset. 

\begin{table}[!ht]
\begin{adjustwidth}{-2in}{0in} 
\centering
\caption{
{\bf Summary information regarding the countries considered for subnational modelling.} The name, number of subnational administration level 1 (admin-1) regions, the year of the first survey collected, the year of the most recently collected DHS survey, and the total number of DHS surveys are given for each country in the dataset.}
\begin{tabular}{ccccc}
\textbf{Country} & \textbf{Total subnational regions} & \textbf{Earliest survey} & \textbf{Latest survey} & \textbf{Total surveys} \\ \thickhline
Afghanistan               & 29 & 2015 & 2015 & 1 \\
Benin                     & 6  & 2001 & 2017 & 4 \\
Burkina Faso              & 14 & 1993 & 2010 & 4 \\
Cameroon                  & 3  & 1991 & 2018 & 4 \\
Cote d'Ivoire             & 15 & 1994 & 2011 & 3 \\
Ethiopia                  & 10 & 2000 & 2019 & 5 \\
Ghana                     & 8  & 1993 & 2014 & 5 \\
Guinea                    & 3  & 1999 & 2018 & 4 \\
Kenya                     & 7  & 1993 & 2014 & 5 \\
Liberia                   & 5  & 2007 & 2019 & 4 \\
Madagascar                & 6  & 1992 & 2021 & 5 \\
Malawi                    & 3  & 1992 & 2016 & 5 \\
Mali                      & 4  & 1995 & 2018 & 5 \\
Mozambique                & 11 & 1997 & 2011 & 3 \\
Myanmar                   & 15 & 2015 & 2015 & 1 \\
Nepal                     & 5  & 1996 & 2016 & 5 \\
Niger                     & 6  & 1992 & 2012 & 4 \\
Nigeria                   & 7  & 1990 & 2018 & 6 \\
Pakistan                  & 6  & 1991 & 2017 & 4 \\
Rwanda                    & 5  & 1992 & 2019 & 7 \\
Senegal                   & 4  & 1992 & 2017 & 9 \\
Tanzania                  & 6  & 1991 & 2015 & 6 \\
Uganda                    & 4  & 1995 & 2016 & 4 \\
Zimbabwe                  & 10 & 1994 & 2015 & 5
\end{tabular}
\label{table1}
\end{adjustwidth}
\end{table}

\section*{Methods}
\subsection*{Modelling approach}
The outcome of interest is the components of a compositional vector, 
\begin{align}
 & \boldsymbol{{\phi_{p,t,m}}}  =(\phi_{p,t,m,1},\phi_{p,t,m,2}) 
\end{align}
where $\phi_{p,t,m,s}$ is the proportion supplied by the public (s =1) or the private (s=2) sector of modern contraceptive method \textit{m}, at time \textit{t}, in subnational administration region \textit{p} and $\sum_{s=1}^{2} \phi_{p,t,m,s}  = 1$. \newline
\newline
The documentation of the modelling assumptions is described using the standarized Temporal Models for Multiple Populations (TMMPs) framework \cite{Susmann2022}. First, we describe the process model which captures the latent dynamics of the outcome of interest. Secondly, we discuss the data model which links the observed data to the process model. In the process model, we model the logit-transformed proportion of the public sector supply share using a Bayesian hierarchical penalised spline model. The logit-transformed data is linked to the process in the data model via a Normal distribution. \newline
Assuming a hierarchical estimation approach accounts for the spatial structure of the data, enabling subnational-level intercepts to benefit from information sharing within each country. Extending this assumption to a multivariate framework across methods allows the model estimates to further leverage latent cross-method correlations embedded in the covariance structure. A key advantage of this approach is the collective information sharing that occurs both across methods and within countries, spanning subnational administrative regions. This mitigates the effects of data sparsity observed in certain subnational regions and countries. Moreover, the approach builds on previous work that forecasts a steady state beyond the most recent survey observation \cite{Comiskey2024} \cite{Track2020}.

\subsubsection*{The process model}
\begin{figure}[!h]
\includegraphics[width=13cm]{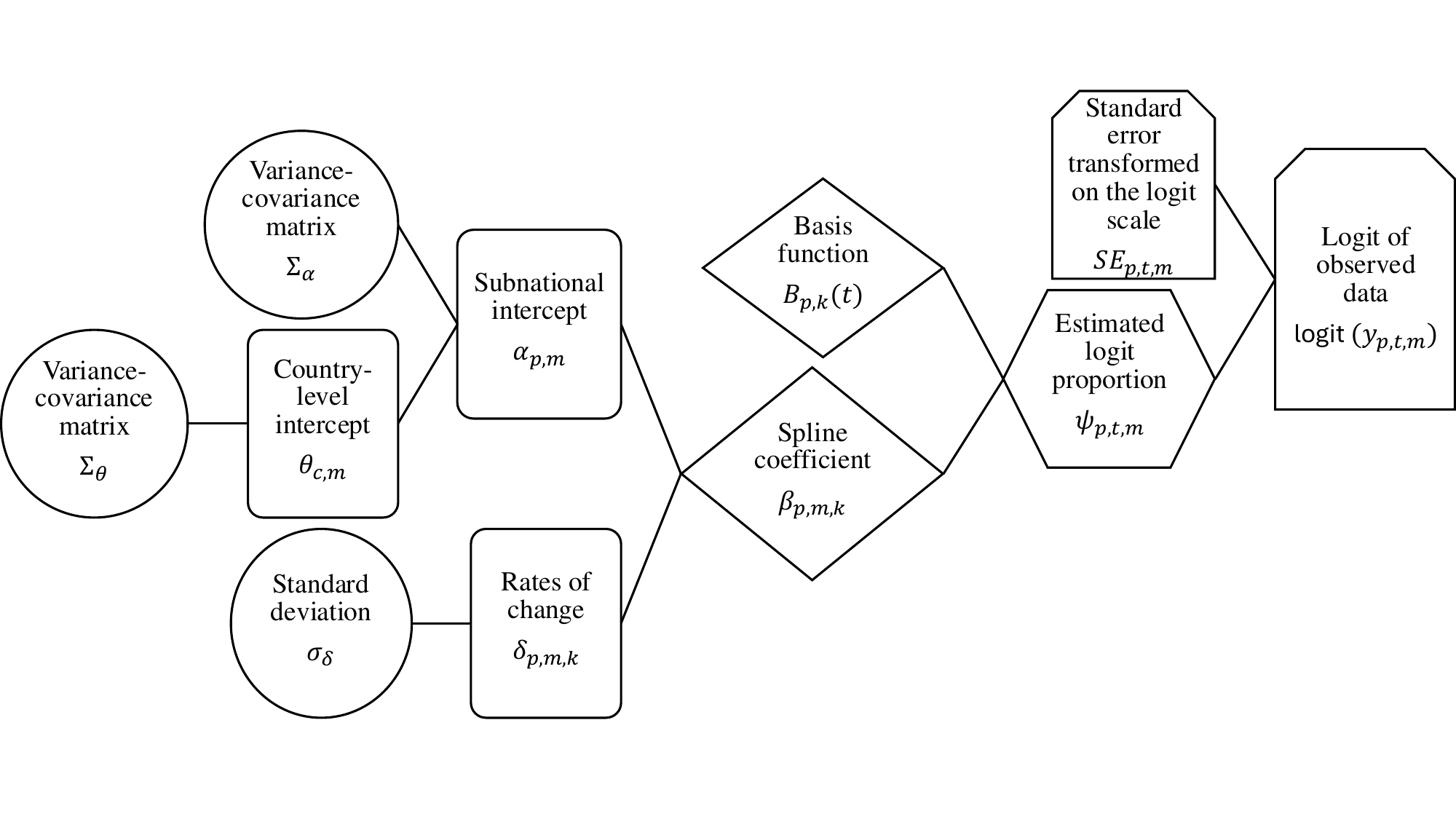}
\caption{{\bf Flow diagram of the subnational supply share model.}
A flow diagram depicting the relationship between all the parameters contributing to the estimation of subnational supply shares. Beginning on the left, the circle boxes depict the variance terms which flow into the second-level of parameters, the intercepts and rates of change (shown in square box with rounded corners). These second-level parameters produce the spline coefficients (shown in diamond box), which in turn combine with the basis functions to form the estimated proportions (shown in hexagonal box). The estimated proportions are linked with the associated standard error calculated using the DHS survey data and transformed onto the logit scale via the Delta method. Together, they link to the observed data (shown in square box with clipped edges). }
\label{fig0}
\end{figure}

The logit-transformed public sector proportion, logit($\phi_{p,t,m,1}$), is modelled via a hierarchical structure depicted in Figure \ref{fig0}. We construct the spline using,

\begin{eqnarray}
\label{eq:eq1}
  \operatorname{logit}\left(\phi_{p,t,m,1}\right) = \psi_{p,t,m} = \sum_{k=1}^{K} \beta_{p,m,k} B_{p,k}(t)
\end{eqnarray}
\newline
where, $\psi_{p,t,m}$ is the latent variable capturing the logit-transformed public sector proportion. $B_{p,k}(t)$ is the k\textsuperscript{th} basis function evaluated at time \textit{t}, in subnational administration region \textit{p}. K is the total number of knots chosen for the set of basis functions. $\beta_{p,m,k}$ is the k\textsuperscript{th} spline coefficient for method \textit{m}, in subnational administration region \textit{p}. Within basis functions, knots are the points where the piece-wise polynomials join together over the input variable (e.g., x-axis). Following the approach outlined in Comiskey et al. (2024), we align a knot with the year of the most recently observed survey in subnational administration region \textit{p}. This results in each subnational administration region \textit{p} having a unique set of basis functions. \newline
\newline
Aligning the knots of the basis function with the most recently observed survey year serves two purposes. Firstly, it allows steady state projections beyond the most recent survey year and secondly, it allows for a forward/backward estimation approach to the vector of K spline coefficients, $\boldsymbol{\beta}_{p,m}$. Let k\textsuperscript{*} be the knot that aligns with the year of the most recently observed survey, then
\begin{eqnarray}
\label{eq:eq4}
\beta_{p, m, k}=\left\{\begin{array}{c}
\alpha_{p, m} \quad k=k^*, \\
\beta_{p, m, k+1}-\delta_{p, m, k} \quad k<k^*, \\ 
\beta_{p, m, k-1}+\delta_{p, m, k-1} \quad k>k^*.
\end{array}\right.
\end{eqnarray}
where,
\begin{eqnarray}
\label{eq:eq3}
\boldsymbol{\delta_{p,m}} = (\beta_{p,m,2} - \beta_{p,m,1}, \beta_{p,m,3}-\beta_{p,m,2}, \dots, \beta_{p,m,K}-\beta_{p,m,K-1}).
\end{eqnarray}
We assume that the value of the spline coefficient at the point k\textsuperscript{*} is $\alpha_{p, m}$. Moving backwards from k\textsuperscript{*} to k=1, we subtract the corresponding rates of change $\delta_{p,m,k}$. Similarly, moving from k\textsuperscript{*} towards k=K, we add the corresponding rates of change $\delta_{p,m,k-1}$. We set a sum-to-zero constraint on the spline coefficients to ensure identifiability of the parameter estimates. The spline coefficient $\alpha_{p, m}$ acts as a proxy intercept in the modelling approach, as a change in $\alpha_{p, m}$ systemically shifts the set of public sector share projections. We assume a hierarchical prior structure using multivariate Normal distributions, such that,
\begin{eqnarray}
\label{eq:eq5}
\theta_{c, 1:M} \mid \Sigma_{\theta}  \sim MVN(\boldsymbol{0}, \Sigma_{\theta}), \\
\alpha_{p, 1:M} \mid \theta_{c, 1:M}, \Sigma_{\alpha}  \sim MVN(\theta_{c, 1:M} , \Sigma_{\alpha}).
\end{eqnarray}
where M is the total number of contraceptive methods considered. For country \textit{c} and set of M methods, we assume that the intercepts $\theta_{c, 1:M}$ are centered on 0 with an associated variance-covariance matrix $\Sigma_{\theta}$. $\theta_{c, 1:M}$  capture the most recently observed publicly supplied proportions at the national level and on the logit scale. $\Sigma_{\theta}$ is MxM matrix which is constant across all countries. Similarly, the M intercepts of subnational administration region \textit{p}, $\alpha_{p, 1:M}$, are centered on a set of country-specific means, $\theta_{c, 1:M}$, with an associated cross-method variance-covariance matrix $\Sigma_{\alpha}$ that is constant across all subnational administration regions. Finally, the variance-covariance matrices assume vague inverse-Wishart priors, centered on an identity matrix of size MxM with M+1 degrees of freedom. The use of this prior assumes that the marginal distributions of the correlations between methods follow a uniform distribution. It is also conjugate prior to the multivariate Normal distribution, making it a suitable choice in our hierarchical estimation process \cite{Zhang2021}.
\begin{eqnarray}
\label{eq:eq6}
\Sigma_{\theta} \sim IW(I^{M}, M+1), \\
\Sigma_{\alpha} \sim IW(I^{M}, M+1).
\end{eqnarray}
The rates of change between spline coefficients, $\delta_{p, m, k}$, are modelled assuming a Normal prior with cross-method variance. We penalise the spline coefficients to remain steady over time, by assuming the expected rate of change to be 0. This acts as a penalty for the B-splines. A vague truncated-Normal prior was used to capture this variation.
\begin{eqnarray}
\label{eq:eq7}
\delta_{p, m, k} \mid \sigma_{\delta} \sim N(0, \sigma_{\delta}) \\
\sigma_{\delta} \sim N(0, 2^{2})_{+}.
\end{eqnarray}

\subsubsection*{The data model}
We link the latent variable $\psi_{p,t,m}$ to the logit-transformed observed public sector supply share, logit(y\textsubscript{p,t,m}), by assuming a Normal distribution likelihood such that, 
\begin{eqnarray}
\label{eq:eq8}
\operatorname{logit}(y_{p,t,m}) \mid \psi_{p,t,m}  \sim N(\psi_{p,t,m}, SE_{p,t,m}^2).
\end{eqnarray}
\newline
Where SE\textsubscript{p,t,m} is associated standard error calculated using the DHS survey microdata and transformed onto the logit scale using the Delta method \cite{Oehlert1992}.

\subsection*{Computation}
We used R and JAGS (Just Another Gibbs Sampler) to fit the model. JAGS uses Markov Chain Monte Carlo (MCMC) algorithm and Gibbs Sampling to produce model estimates for Bayesian Hierarchical models \cite{PlummerJAGS}. To evaluate the JAGS output we used `rjags', an R package that offers cross-platform support from JAGS to the R interface \cite{rjags}. The number of iterations used was 80,000. The burn-in period was set to 10,000. The samples were thinned to every 35th sample. Consequently, the posterior distribution is made up of 2000 samples. To assess convergence, we considered the R-hat values of the model parameters using the plot function of rjags, as well as the trace plots and autocorrelation function plots of individual parameters \cite{Vehtari2021}. The results were a set of trajectories for the proportion of contraceptive \textit{m} supplied by the public and private sectors over time for each subnational region included in the study. The median of these results was taken to be the model's point estimate. The 95\% credible intervals were calculated using the 2.5th and 97.5th percentiles from the posterior distribution for each estimate. 

\section*{Results}
\subsection*{National and subnational correlations}
\begin{figure}[!h]
\includegraphics[width=13cm]{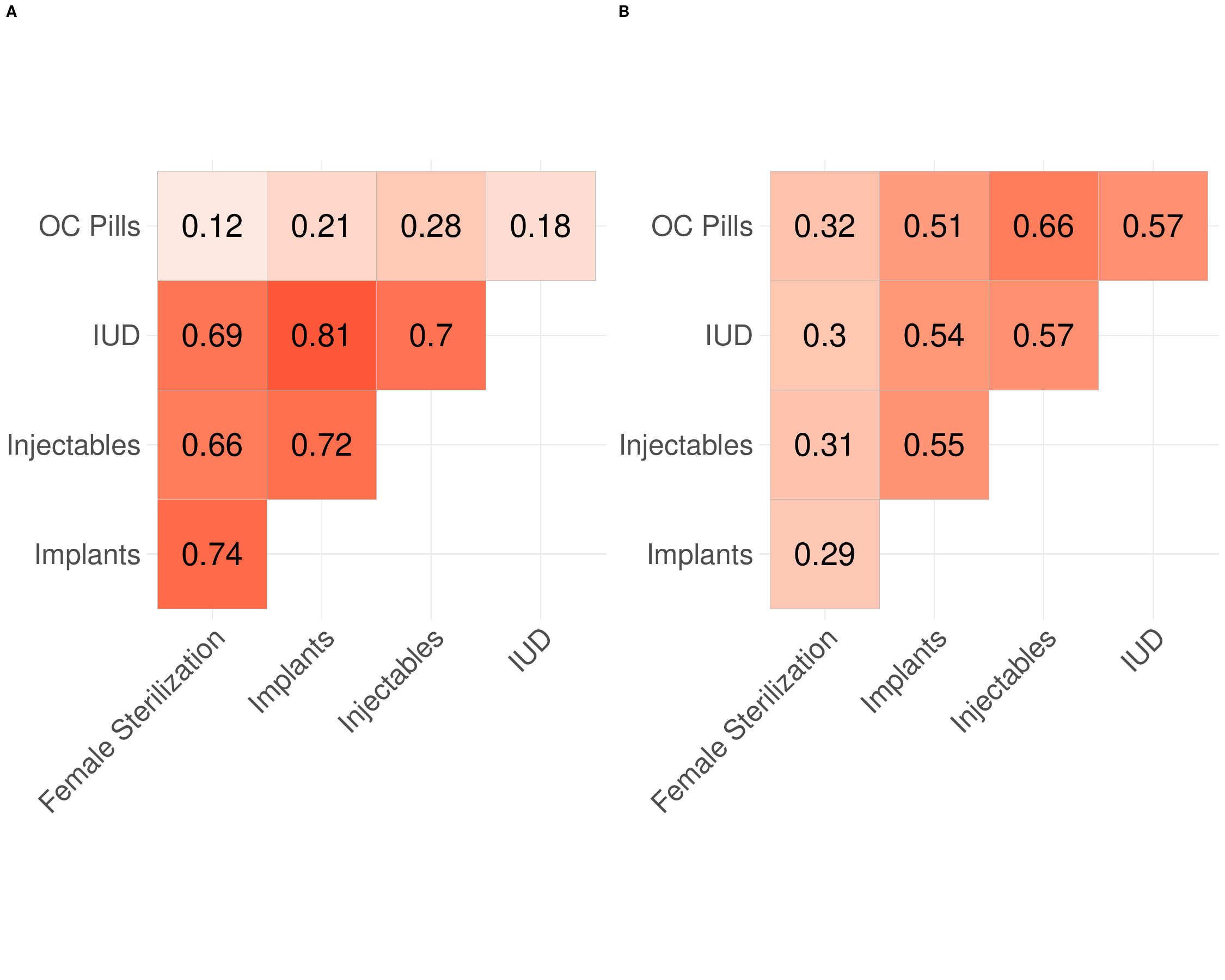}
\caption{{\bf Estimated (A) national- and (B) subnational-level cross-method correlations informing $\alpha_{p, m}$, the proxy-intercept.} The estimated correlations captured by the Wishart prior for the intercept term of the proposed model. The intercept is informed by the most recently observed public supply shares of each method across all (A) countries and (B) subnational administration regions. Each of the five methods is listed along the x- and y-axes, with the estimated correlation is given in each square. The strength of the correlation is emphasised by the depth of the shade. Lighter infer correlations closer to 0, while darker colours infer stronger correlations.}
\label{fig1}
\end{figure}

Fig. \ref{fig1} shows the heat map of the estimated latent correlations between the five contraceptive methods considered in this study informing $\alpha_{p, m}$, the proxy-intercept, at the national (A) and subnational (B) levels across all subnational provinces. At the national level (Fig. \ref{fig1}A), the estimated correlations are showing strong positive relationships between the long-term and permanent methods (LAPM). For OC pills, a short-term method, the estimated correlations with the other methods are weakly positive. This trend is in keeping with previous studies that found OC pills are a popular choice for women accessing contraceptive through the private sector \cite{Corroon2016} \cite{Weinberger2017} \cite{Bradley2022}. In contrast, contraceptive methods that require a medical professional for administration are more likely to be accessed through the public sector. These correlations would imply that as the most recently observed public sector supply share of a given LAPM increases, the other LAPMs also tend to increase. For the subnational-level correlations (Fig. \ref{fig1}B), the estimated correlations between female sterilisation and the other methods are the weakest. The remaining cross-method correlations are moderately strong between 0.51 (OC pills and implants) and 0.66 (OC pills and injectables). The estimated cross-method correlations at the subnational level are more homogenous than those observed at the national level. This may be due in part to the noisier signal observed at the subnational level.

\subsection*{Country case studies}
 While we produced model estimates based on model fit to the subnational provinces of 24 countries, we have chosen two case-study countries Nigeria and Rwanda to showcase the strength of our modelling approach. Both countries have varying amounts of data across the five contraceptive methods and have experienced changes in the modern contraceptive supply share trends over time. 

\subsubsection*{Nigeria}
\begin{figure}[!h]
\includegraphics[width=13cm]{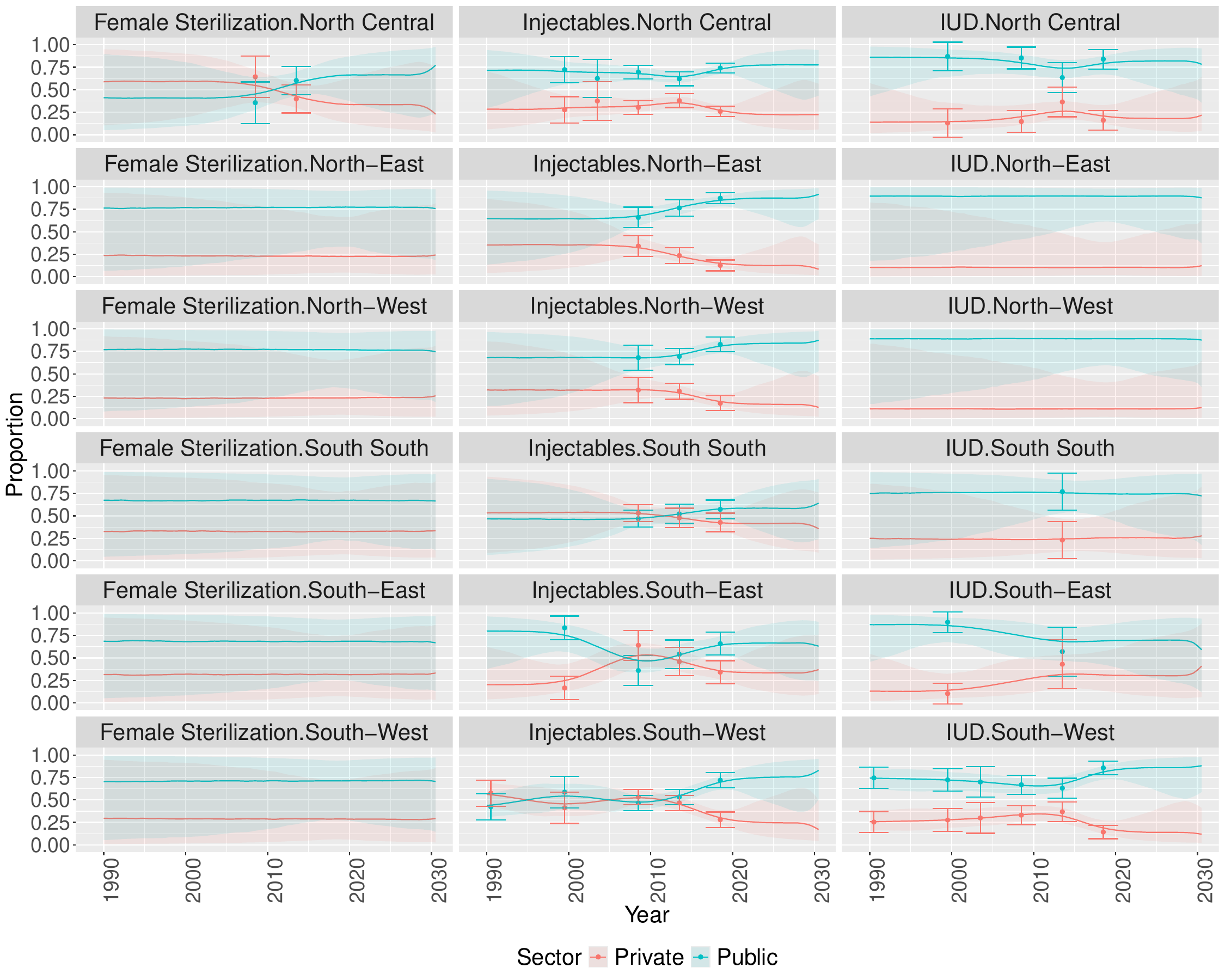}
\caption{{\bf Estimated subnational contraceptive supply shares in Nigeria with uncertainty.}
The projections for the proportion of modern contraceptives supplied by each sector for three of the five contraceptive methods, in the six geopolitical zones of Nigeria. The median estimates are shown by the continuous line while the 95\% credible interval is marked by shaded coloured areas. The DHS data point is signified by a point on the graph with error bars displaying the standard error associated with each observation. The sectors are shown by blue triangles for public and red for the private sector.}
\label{fig2}
\end{figure}

\begin{figure}[!h]
\includegraphics[width=13cm]{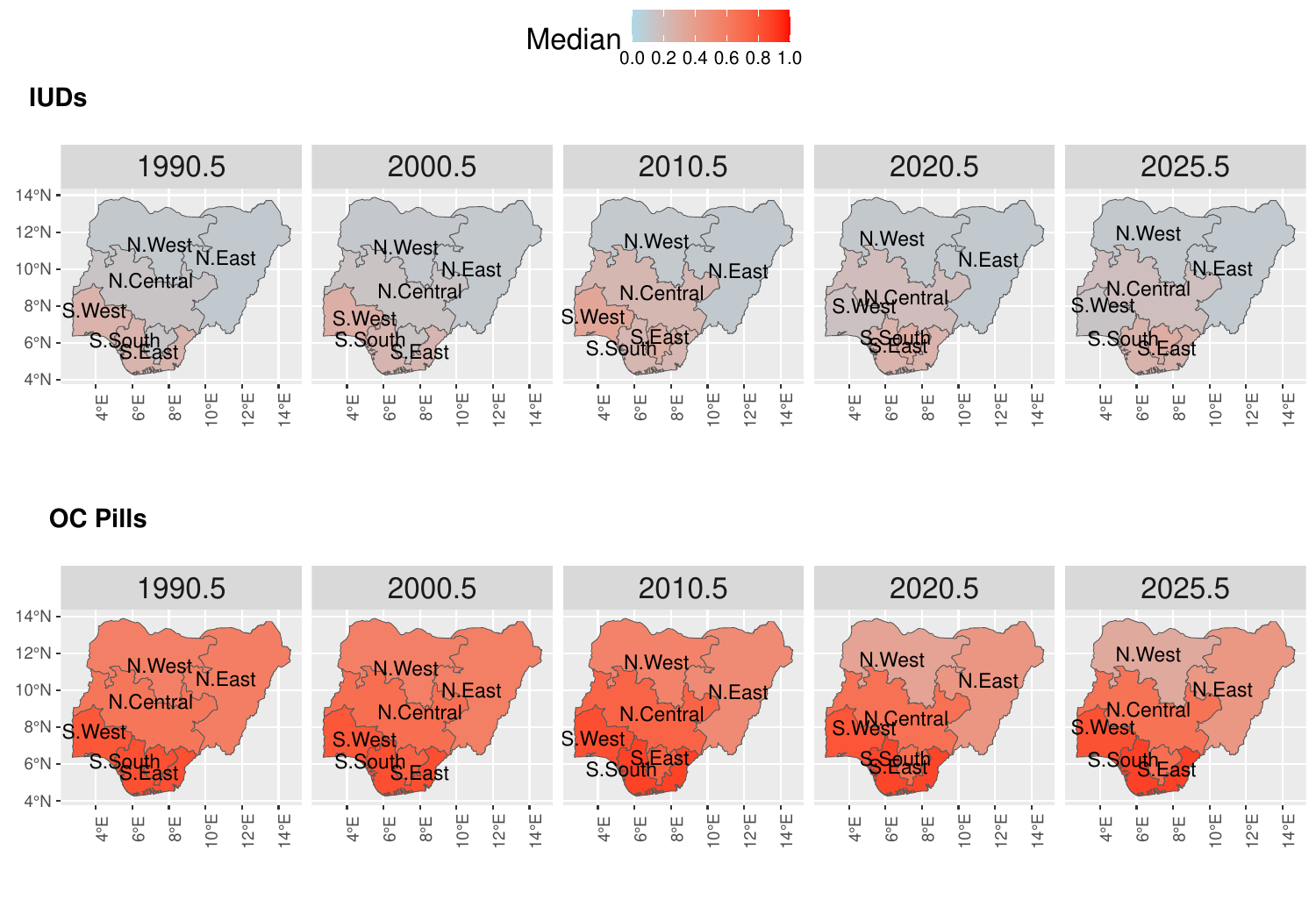}
\caption{{\bf Map of estimated subnational contraceptive private sector supply shares in Nigeria over time.} The median estimates of private sector contraceptive method supply shares for two of the five contraceptive methods, in the six geopolitical zones of Nigeria. The colour captures the median supply shares where light blue colours capture supply shares near 0 while the deep red colour captures supply shares approaching 100\%. Five years of estimates are given at 10-yearly intervals from 1990 to 2020 and then the present year, 2025. Label `IUDs' shows the supply of intra-uterine devices. Label `OC Pills' shows the supply of oral contraceptive pills. }
\label{fig3}
\end{figure}

Fig. \ref{fig2} shows the estimated proportion of contraceptives supplied by the public and private sectors with uncertainty from 1990 to 2030 across the six geopolitical zones. The boundaries of these zones have been consistent for the duration of the observation period. Historically, only the North Central region of Nigeria has collected data on the supply of female sterilization. As such, the remaining regions are estimated via the hierarchical estimation process on the intercept. The estimated method supply share in the remaining regions is given by the most recently supplied share in the North Central region. With no data to inform any deviation from this observed level, coupled with the splines being penalised to project steadily into the future, we see flat trends occurring for these regions. In the North Central region, the model indicates an increasing trend between 2015 and 2020, approximately. In contrast to the data sparsity seen in female sterilization, injectables have between three and five survey observations per region. The flexibility of the splines is demonstrated here as the model is able to capture the complex nature of the data in each region. Historically, OC pills tends mostly to be provided by the private sector across all subnational regions. However, the model estimates indicate that this private sector supply share has been steadily declining over time. This is in contrast to the supply trends observed in other methods, where the public sector is the main supplier across all regions, and has remained relatively constant over time. Some smoothing is taking place as the data model is informed by the observed standard errors of the survey data. For this reason, observations with large associated standard errors are not as closely adhered to, when compared to observations with smaller uncertainty. This is clearly seen in North-Central OC pills where the model estimates smooth through the survey observation at 2003. 
When we consider the private sector supply shares spatially and over time, we can see geographical trends in the supply of certain methods (Fig. \ref{fig3}). For instance with IUDs (Fig. \ref{fig3}: IUDs), there appears to be a north-south split across Nigeria in the private sector supply over time. The northern half (North East and North West subnational regions) tend towards very low proportions of IUDs provided by the private sector while the southern half of the country (North Central, South West, South South, South East) sees higher proportions of IUDs supplied by the private sector. This trend is replicated in OC pills (Fig. \ref{fig3}: OC Pills) where again, the northern half of the country tends to supply a lower proportion of OC pills via the private sector when compared to the southern half. 

\subsubsection*{Rwanda}

In Rwanda, we estimate contraceptive method supply shares for the five provinces \ref{fig2}. The boundaries of these zones have been consistent for the duration of the observation period. Historically, the public sector has dominated the supply of all five contraceptive methods across all five provinces since 1990. The supply of implants is almost exclusively provided by the public sector, with very small proportions across all methods being privately sourced. The supply of IUDs in Kigali is at an approximately 60:40 split between the public and private sector. This is quite different to the trends we observe in the other regions where the public sector takes over 80\% of the market share. The contraceptive supply market has been turbulent in Rwanda across all the regions for injectables and OC pills. In injectables, from 1990 to 2015 the market shares had remained very steady over time with little private sector involvement. However, we see that post-2015 the private sector rapidly increases it's share of the injectables supply market and post-2020 it overtakes the public sector as the largest supplier of injectables in all regions. Similarly, in OC pills we see a rapid increase of private sector supply shares post-2015. In Kagali and the North we see that the private sector is estimated to become the largest supplier of OC pills post-2019. In the South, the OC pill market share is split at an approximate 50:50 between public and private shares. Figure \ref{fig5} captures the median trends of the private sector supply shares spatially and temporally. Across all methods, we see that Kigali provinces tends to have the strongest private sector supply share over time. From Figure \ref{fig5}, we see that Kigali tends to increase the private sector supply share of a given method and then the neighboring regions follow suit afterwards. Overall, private sector supply shares have historically been low in Rwanda across all regions, with the exception of OC pills (Figure \ref{fig5}E) and injectables (Figure \ref{fig5}C) in recent years. 

\begin{figure}[!h]
\includegraphics[width=13cm]{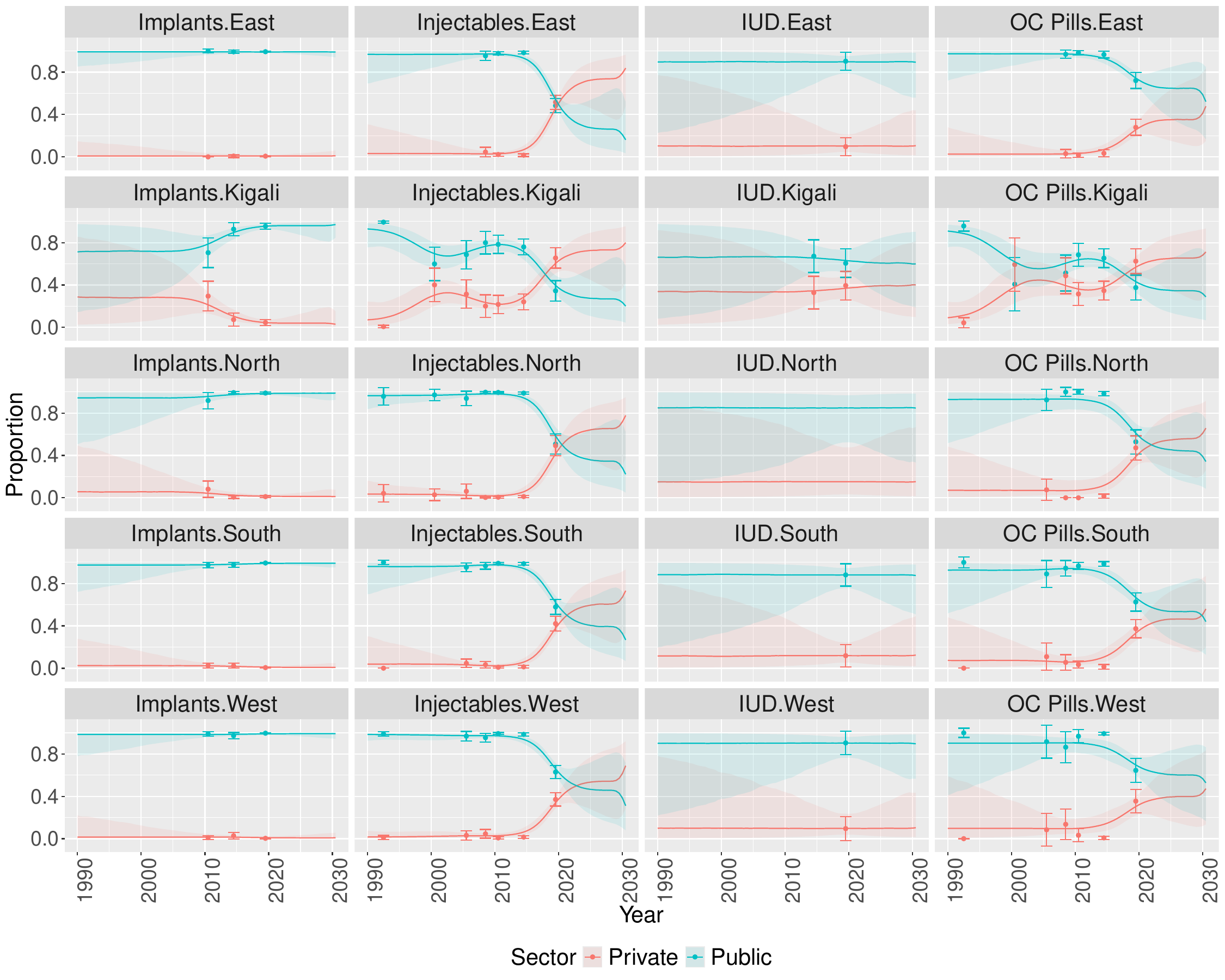}
\caption{{\bf Estimated subnational contraceptive supply shares in Rwanda with uncertainty.}
The projections for the proportion of modern contraceptives supplied by each sector for four of the five contraceptive methods, in the five provinces of Nigeria. The median estimates are shown by the continuous line while the 95\% credible interval is marked by shaded coloured areas. The DHS data point is signified by a point on the graph with error bars displaying the standard error associated with each observation. The sectors are shown by blue triangles for public and red for the private sector.}
\label{fig4}
\end{figure}

\begin{figure}[!h]
\includegraphics[width=13cm]{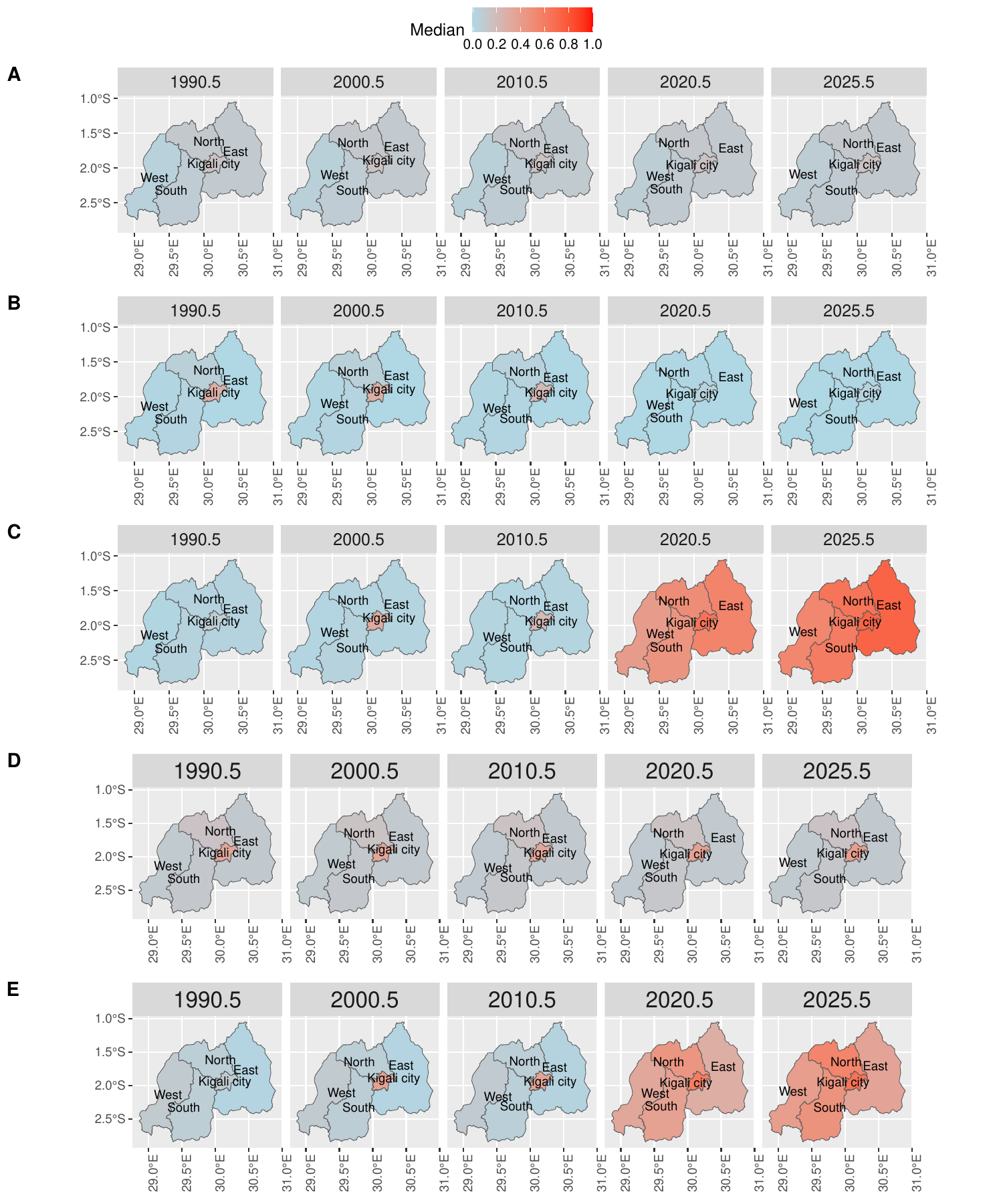}
\caption{{\bf Map of estimated subnational contraceptive supply shares in Rwanda over time.} The median estimates of private sector supply shares of the five contraceptive methods across the five provinces of Nigeria. The colour captures the median supply shares where light blue colours capture supply shares near 0 while the deep red colour captures supply shares approaching 100\%. Five years of estimates are given at 10-yearly intervals from 1990 to 2020 and then the present year, 2025. The A label shows the supply of female sterilisation. The B label shows the supply of implants. Label C shows the supply of injectables. Label D shows the supply of intra-uterine devices (IUDs). Label E shows the supply of oral contraceptive pills. }
\label{fig5}
\end{figure}

\subsection*{2025 estimate summary}

 In Table \ref{table2}, we present the estimated median method supply shares across all subnational administration regions for the year 2025. Overall, the public sector dominates in the supply of modern contraceptives at the subnational level. Implants are most commonly accessed through the public sector with a median of almost 88 percentage points. OC pills show an approximate 50:50 split between public and private sector market shares. In comparison to OC pills, a short-term contraceptive method, the long-acting and permanent methods, female sterilization, implants, injectables, and IUDs, have higher public sector supply shares by at least 20 percentage points in each instance. This is consistent with previous research where women were found to be more likely to access short-term methods via private facilities \cite{Corroon2016} \cite{Weinberger2017} \cite{Bradley2022}. These findings are also consistent with the contraceptive method supply shares estimated at the national-level in 2023 \cite{Comiskey2024}.

\begin{table}[!ht]
\centering
\caption{
{\bf Summary of estimated contraceptive supply shares across all subnational administration regions in 2025 for each method and sector.} The median estimate for the proportion supplied and the associated standard deviation (SD) are listed.}
\begin{tabular}{|c|cc|cc|}
\hline
                     & \multicolumn{2}{c|}{\textbf{Public}} & \multicolumn{2}{c|}{\textbf{Private}} \\ \hline
                              & \multicolumn{1}{c|}{\textbf{Median}} & \textbf{SD} & \multicolumn{1}{c|}{\textbf{Median}} & \textbf{SD} \\ \hline
\textbf{Female Sterilization} & \multicolumn{1}{c|}{0.792}           & 0.104       & \multicolumn{1}{c|}{0.207}           & 0.104       \\ \hline
\textbf{Implants}    & \multicolumn{1}{c|}{0.881}  & 0.160  & \multicolumn{1}{c|}{0.119}   & 0.157  \\ \hline
\textbf{Injectables} & \multicolumn{1}{c|}{0.823}  & 0.173  & \multicolumn{1}{c|}{0.177}   & 0.173  \\ \hline
\textbf{IUD}         & \multicolumn{1}{c|}{0.822}  & 0.142  & \multicolumn{1}{c|}{0.178}   & 0.142  \\ \hline
\textbf{OC Pills}    & \multicolumn{1}{c|}{0.512}  & 0.262  & \multicolumn{1}{c|}{0.487}   & 0.262  \\ \hline
\end{tabular}
\label{table2}
\end{table}

\subsection*{Validation results}
The model described in this paper has been validated using several metrics to assess its effectiveness at estimating subnational method supply shares over time. We split our complete data into testing and training datasets, any data beyond 2015 was labeled as the test set and withheld for validation. Further details of the validation approach and the metrics used to evaluate it's performance can be found in the Supporting Information. Given the complex nature of the data, the model is performing reasonably well. Table \ref{S_tab1} gives the validation results for the Multivariate intercept P-spline model, described in the Methods section, and the modelling alternatives, described in the Supporting Information. First we consider the mean absolute relative error (MARE), a negatively orientated accuracy measure of each approach using the test set. The Multivariate intercept P-spline model has the lowest MARE at approximately 7 percentage points while the 0-covariance P-spline model saw the largest MARE at approximately 12 percentage points. We also consider the standardized absolute prediction error (SAPE), which is a measure of the dispersion of the generated predictive distributions for the test set.  In all modelling cases, the SAPE was greater than 1. This would imply that the data is more spread out that the models are predicting. The lowest SAPE was recorded with the Fully Multivariate P-spline model at 2.81 while the largest SAPE was recorded for the shrinkage P-spline model. Our proposed model had the second smallest SAPE at 3.46. For coverage, in the 80\% instance we would expect our models to capture 80\% of the test set. There was some over-fitting for this metric, with the Multivariate intercept P-spline model giving a coverage of approximately 82\%. This was the closest coverage to the expected 80\% of all 5 models. The shrinkage P-spline model had the lowest coverage at approximately 77\%. In the 95\% case, we expect our models to capture 95\% of the test set. In this instance, we do not see the same over-fitting issue as the 80\% coverage metric. The fully multivariate P-spline model scores slightly higher than the proposed model in this instance with an approximate coverage of 95\%. However, the proposed model has a marginally smaller coverage of approximately 94 percentage points. To evaluate the accuracy of our model, we considered the root mean square error (RMSE) of the public sector test set estimates. The lowest RMSE is observed in our proposed model with an approximately average error of 15 percentage points. The highest RMSE was observed in the Multivariate delta P-spline model at approximately 19 percentage points. The width of the prediction intervals (PI width) is a crucial validation metric because it directly impacts the usability of the estimates. If the PI width is too narrow, the intervals may fail to capture the true values, leading to poor coverage. On the other hand, if the PI width is too wide, the estimates become less informative and lose their practical value in other modelling scenarios. The smallest median 95\% PI width was observed in the shrinkage P-spline model at approximately 44 percentage points. The largest PI width was observed for the Multivariate delta P-spline model at approximately 53 percentage points. Notably, the Multivariate intercept P-spline model provides the second smallest PI width at approximately 47 percentage points. This is 5 percentage points lower than the Multivariate delta P-spline model. Finally, in an unbiased model we would expect to see an even split of the incorrectly estimated test set observations above and below the prediction interval bounds. In all modelling instances, there are a higher proportion of incorrectly estimated observations below the 95\% prediction interval. This would imply that the models tend to over-estimate the data. Given these results, we conclude that the proposed Multivariate intercept P-spline model is the performs the best overall. It has good coverage in both the 80\% and 95\% instances. The width of the prediction interval is not as large as other modelling alternatives, indicating that this updated model has lower uncertainty in the model estimates compared to the others. In addition, it has the lowest RMSE and MARE, and the second lowest SAPE value of all the models. Overall, the method seems to have provided reasonably accurate and well-calibrated probabilistic projections for the 2015–2022 period.

\begin{table}[!ht]
\begin{adjustwidth}{-2.25in}{0in} 
\centering
\caption{
{\bf Model validation results.} The validation results for the test set across different potential modelling approaches. MARE is mean absolute relative error and is a percentage. SAPE is the standardized absolute prediction error and is measured on the logit scale. Coverage is the proportion of the test set observations that are captured within the 80\% and 95\% prediction intervals produced by the model. RMSE is root mean square error.  We consider the median PI width and evaluate the location of the incorrectly estimated leave-one-out validation test set observations. The Multivariate intercept P-spline model is described in the Methods section, while the modelling alternatives are described in the Supplementary Materials.}
\begin{tabular}{|c|c|c|c|c|c|c|}
\hline
\textbf{Metric} &
   &
  \textbf{\begin{tabular}[c]{@{}c@{}}Multivariate \\ intercept \\ P-spline model\end{tabular}} &
  \textbf{\begin{tabular}[c]{@{}c@{}}Multivariate \\ delta \\ P-spline model\end{tabular}} &
  \textbf{\begin{tabular}[c]{@{}c@{}}0-covariance \\ P-spline model\end{tabular}} &
  \textbf{\begin{tabular}[c]{@{}c@{}}Shrinkage \\ P-spline model\end{tabular}} &
  \textbf{\begin{tabular}[c]{@{}c@{}}Fully \\ multivariate \\ P-spline model\end{tabular}} \\ \hline
\textbf{\begin{tabular}[c]{@{}c@{}}MARE \\ (\%)\end{tabular}} &
   &
  7.42 &
  11.25 &
  11.86 &
  11.16 &
  9.10 \\ \hline
\textbf{SAPE} &
   &
  3.46 &
  2.94 &
  3.55 &
  3.80 &
  2.81 \\ \hline
\textbf{\begin{tabular}[c]{@{}c@{}}80\% \\ coverage \\ (\%)\end{tabular}} &
   &
  81.95 &
  82.54 &
  77.51 &
  76.62 &
  82.54 \\ \hline
\textbf{\begin{tabular}[c]{@{}c@{}}95\% \\ coverage \\ (\%)\end{tabular}} &
   &
  93.19 &
  92.01 &
  91.42 &
  90.82 &
  93.49 \\ \hline
\textbf{\begin{tabular}[c]{@{}c@{}}RMSE \\ (\%)\end{tabular}} &
   &
  15.26 &
  19.39 &
  17.97 &
  18.85 &
  16.19 \\ \hline
\textbf{\begin{tabular}[c]{@{}c@{}}Median 95\% \\ PI width (\%)\end{tabular}} &
   &
  46.50 &
  52.60 &
  47.40 &
  44.1 &
  50.9 \\ \hline
\multirow{2}{*}{\textbf{\begin{tabular}[c]{@{}c@{}}Location of \\ incorrectly \\ estimated \\ observations\end{tabular}}} &
  \textit{\begin{tabular}[c]{@{}c@{}}Above \\ the \\ 95\% PI\end{tabular}} &
  2.37 &
  2.07 &
  3.55 &
  3.55 &
  1.48 \\ \cline{2-7} 
 &
  \textit{\begin{tabular}[c]{@{}c@{}}Below\\ the \\ 95\% PI \end{tabular}} &
  4.44 &
  5.92 &
  4.73 &
  5.62 &
  5.03 \\ \hline
\end{tabular}
\label{S_tab1}
\end{adjustwidth}
\end{table}

\section*{Discussion}
In this paper, we have developed a Bayesian method for probabilistic projections of subnational contraceptive supply shares over time with available DHS survey data. The modelling framework is based on Bayesian hierarchical models and uses penalized splines to capture the evolution of the contraceptive method supply from the public and private sectors while imposing a correlation structure to capture correlations between the most recently observed public sector supply shares. The hierarchical nature of the modelling framework allows for cross-subnational information sharing within countries to promote precise estimation, even when limited data is available. The posterior predictive distributions are estimated using past DHS data for all subnational administration regions from 1990 to 2022. The model will produce estimates within the period of available survey data as well as projections beyond the most recent data point. The resulting predictive distributions were accurate and reasonably well calibrated in an out of sample validation exercise for forecasting the most recent seven year period. The modelling framework was found to outperform other suitable alternative modelling approaches. \newline
\newline
This modelling framework has several advantages over the aforementioned modelling alternatives, and makes another contribution to the estimation of subnational proportions. Firstly, using the Bayesian hierarchical framework accounts for the spatial nature of the data while also allowing subnational regions within a country to share information. An advantage of this is that information sharing within countries improves the precision of the resulting subnational estimates and informs model estimates where previously no data was present, all in a data-driven approach. Secondly, using a multivariate approach to hierarchal modelling accounts for the correlations that exist between method supply shares, and allows for another information exchange to occur, this time across contraceptive methods. Estimating these cross-method national- and subnational-level correlations further explains the latent trends underpinning the complexity nature of subnational method supply shares. Incorporating the observed standard errors of the DHS estimates into the data model further controls the uncertainty associated with the resulting estimates, and promotes the smooth model estimates and projections from an otherwise heterogeneous dataset. Finally, using penalised splines allows for data driven, flexible model-based estimates. For these reasons, we believe that proposed model makes a valuable contribution to the area of subnational estimation of proportions, in data sparse settings. \newline
In addition, the approach introduces two key improvements over previous approaches of Comiskey et al. (2023) and Comiskey et al. (2024). First, it captures non-zero covariances between proxy-intercepts rather than between rates of change. Since subnational data tends to be noisier than national-level supply share data, correlations between rates of change are often difficult to estimate and uncertain. By allowing the most recently observed levels to share information across methods, this approach enhances estimate precision in data-sparse regions. Second, the model adopts a fully Bayesian process for estimating covariance matrices, replacing the earlier method of Comiskey et al. (2024) that relied on \textit{a posteriori} correlation estimates. This update is not only more computationally efficient but also avoids restricting the parameter space of the resulting covariance matrix. Overall, this updated model more effectively captures the complexities of subnational data, while still offering insights into cross-method correlations and spatial relationships found in the previous approach of Comiskey et al. (2024). \newline
\newline
In recent years, a substantial number of low- and middle-income countries have implemented a dencentralised health sector making the estimation of health indicators in data-sparse, small-areas of vital importance to researchers and policy-makers \cite{Bossert2002}. However, the lack of reliable data and regular subnational estimates of key indicators have reduced the ability of health and development authorities’ to strengthen the delivery of contraception and other reproductive health services via local systems \cite{Li2019}.  When evaluating health programmes, policymakers must consider cost, access and quality \cite{Shah2011}. A potential application of these subnational method supply shares is in the evaluation of access. Strengthening access to contraceptives through public-private sector collaborations and complementing ongoing public-private partnerships for the advocacy of family planning can lead to an increase in contraceptive uptake \cite{White2016}. It is well-established that teenagers and sexually active unmarried women tend to use private sector suppliers of contraception \cite{Elnakib2022}. However, as seen in our country case studies of subnational contraceptive supplied in Nigeria and Rwanda, many contraceptives are supplied almost solely by the public sector across many subnational regions. These model estimates provide valuable insights for policymakers into the temporal and spatial dynamics of the subnational contraceptive supply share markets, highlighting areas for potential improvement in supply dynamics and equity. Having public and private sectors both supplying contraceptives is key to achieving equity among family planning users \cite{Bradley2022}, but our case studies have shown that for some contraceptives the private sector is supplying very low proportions of some contraceptives over time. Lastly, a total-market approach is essential to the success and sustainability of the family planning market. Strengthening supply chains through informed data-driven supply share estimates will contribute to ensuring contraceptive security for women and girls globally \cite{Moazzam2017}. \newline
\newline
In future work, we hope to incorporate additional demographic and family planning indicator covariates into the modelling framework to further improve the precision of the model estimates. We hope to extend the framework to account for multiple data sources, accounting for the different sources of error.

\section*{Supporting information}

\subsection*{Subnational administration model estimates}
The complete set of subnational supply share estimates for every country and subnational administration region can be reproduced using the code and data found on the github, \url{https://github.com/hannahcomiskey/subnational_estimates/}.

\subsection*{Out of sample validation}
\subsubsection*{Metrics}
To validate our model, we split our data into test and training sets. The observed survey data from 1990 to 2014 inclusive was considered the training set, while any data beyond 2015 was labeled as the test set and withheld for validation. Using the training set, we estimate a predictive distribution of the method supply shares supplied by the public sector for each method and subnational administration region in our training dataset from 1990 to 2024. We then compared the resulting predictive distributions with the observed public sector test set data. The test set contained 16 countries, covering 81 subnational administration regions and all five contraceptive methods. In total, the test set contained 338 observations while the training set contains 931 observations. We evaluated the results of the validation using different measures of accuracy and prediction interval calibration. \newline
\newline 
To assess the accuracy of our mean point predictions, we consider the mean absolute relative error (MARE) in percentage points. The MARE is negatively orientated, therefore smaller MARE values are better. The MARE is given by:
\begin{eqnarray}
\label{eq:S1eq}
  MARE = 100 \times \frac{1}{PTM} \sum_{p,t,m} \frac{|y_{p,t,m} - \hat{y}_{p,t,m}|}{\hat{y}_{p,t,m}}
\end{eqnarray}
where, y\textsubscript{p,t,m} is the observed proportion of method \textit{m}, at time \textit{t}, in subnational administration region \textit{p} supplied by the public sector. y\textsubscript{p,t,m} is part of the test set. $\hat{y}_{p,t,m}$ is the corresponding estimated proportion supplied by the public sector. P is the number of subnational administration regions in the test set. T is the number of time periods in the test set. M is the number of contraceptive methods in the test set. \newline
\newline
The standardized absolute prediction error (SAPE) is a measure of the dispersion of the generated predictive distributions. Ideally, we expect the SAPE value to be 1. When the values observed in the test set are more spread out than the predictive distributions, we say that the distribution is under-dispersed. In these instances the SAPE value will be greater than 1. Conversely, when the predictive distribution is over-dispersed, the SAPE value will be less than 1. The SAPE is given by: 
\begin{eqnarray}
\label{eq:S2eq}
  SAPE = 1.4826 \times \operatorname{median}(\frac{|y_{p,t,m} - \hat{y}_{p,t,m}|}{\hat{\sigma}_{p,t,m}})
\end{eqnarray}
where $\hat{\sigma}_{p,t,m}$ is the estimated Bayesian predictive standard deviation associated with the test set observation y\textsubscript{p,t,m}.\newline
\newline
We calculate error terms, e\textsubscript{p,t,m}, to describe the difference between the test set public sector survey observation, y\textsubscript{p,t,m}, and the corresponding median estimate from the posterior predictive distribution, $\hat{y}_{p,t,m}$ such that,
\begin{eqnarray}
\label{eq:S3eq}
    e_{p,t,m} = y_{p,t,m} - \hat{y}_{p,t,m}.
\end{eqnarray}
To evaluate the accuracy of our model, we considered the root mean square error (RMSE) of the public sector test set estimates. Let,
\begin{eqnarray}
\label{eq:S4eq}
    \text{RMSE}= \sqrt{\frac{\sum_{p,t,m} e_{p,t,m}^2}{N}},
\end{eqnarray}
where, $N$ is the number of observations in the test set. e\textsubscript{p,t,m} is the error calculated for the proportion of method \textit{m}, at time \textit{t}, in subnational administration region \textit{p} supplied by the public sector. The RMSE can be interpreted as the average error observed across all subnational administration regions, time points and methods in the test set.  \newline
\newline
Coverage assumes that if our model is correctly calibrated, then for each sector the model should be able to capture the test set of out-of-sample observations with 80\% and 95\% accuracy.  To examine the bias of our models estimates, we examined the location of the incorrectly estimated test set observations for the 95\% coverage instance. We expect that the 5\% of observatsions are incorrectly estimated. These incorrectly estimated observations should approximately evenly distributed above and below the estimated 95\% prediction interval. By examining the breakdown of location, we are evaluating the tendency of the bias of our model, i.e. whether the model tends to under- or over-estimate the test set. A larger proportion of observations located below the prediction interval, shows that the model is tending to over-estimate the test set. Conversely, when a higher proportion of the incorrectly estimated observations are located above the prediction interval, the model tends to under-estimate the test set. Finally, we also consider the median estimate of the 95\% predictive interval widths for each test set observation across all subnational administration regions, time points and methods.

\subsubsection*{Modelling alternatives}
To justify the complexity of our proposed model, the 'multivariate intercept P-spline model', we compared it against suitable alternatives. To begin, we compared the proposed model against the model described in Comiskey et al. (2024). We refer to this model as the 'multivariate delta' model. We ran the multivariate delta model at the subnational level, including estimated subnational correlations across methods for the rates of change in spline coefficients, for the public/private sector breakdown of method supply shares over time. We also took the 0-covariance model from this paper to include as a baseline model. In this instance, we refer to the 0-covariance model of Comiskey et al. 2024 as the '0-covariance P-spline' model. For complete descriptions of these models, please refer to the aforementioned paper. \newline 
\newline
To evaluate the impact of estimating our spline coefficients using the forward/backward approach, we also include B-spline models with spline coefficients estimated using a traditional AR(1) process. \newline 
\newline
To begin, we consider a hierarchical Bayesian model that combines basis splines for smooth predictions over time, with method-, subnational administration region-specific intercepts and shrinkage triple gamma priors for regularization \cite{Cadonna2020}. We will refer to the model as the 'shrinkage P-spline' model.  As with our proposed model, we set a sum to zero constraint on the spline coefficients to ensure identifiability. The B-spline model links the observed data to the model’s predictions using a logit-normal data model, informed by the observed logit-transformed standard errors of the survey data. For the public sector component of interest, $\phi_{p,t,m,1}$, the B-spline model is set up as follows:
\begin{eqnarray}
  \operatorname{logit}\left(\phi_{p,t,m,1}\right) = \psi_{p,t,m} = \alpha_{p,m} + \sum_{k=1}^{K} \beta_{p,m,k} B_{k}(t)
\label{eq:S5eq}
\end{eqnarray}
Where, the intercept term, $\alpha_{p,m}$, is estimated hierarchically such that,
\begin{eqnarray}
\alpha_{c, m}^{country} \mid \sigma_{\alpha_{m}^{country}}  \sim N(0, \sigma_{\alpha_{m}^{country}}^{2}), \\
\alpha_{p, m}^{prov.} \mid \alpha_{c[p],m}^{country}, \sigma_{\alpha_{m}^{prov.}}  \sim N(\alpha_{c[p],m}^{country} , \sigma_{\alpha_{m}^{prov.}}^{2} ). 
\label{eq:S6eq}
\end{eqnarray}
The standard deviation terms of the hierarchically estimated intercept terms are given vague truncated Cauchy and Normal priors. Such that,
\begin{eqnarray}
\sigma_{\alpha_{m}^{country}} \sim C^{+}(0,1),\\
\sigma_{\alpha_{m}^{prov.}} \sim N^{+}(0,2^{2}).
\label{eq:S7eq}
\end{eqnarray}
To estimate the spline coefficients, we use an auto-regressive process such that the expected value of each spline coefficient is that of the previous spline coefficient. Included in this estimation process is a triple gamma shrinkage prior. This prior shrinks the effects of a given spline coefficient towards 0 when it is not considered a significant contributor \cite{Cadonna2020}. Regularizing the spline coefficients via a shrinkage prior reduces the possibility of over-fitting. For this model, we used the default parameter settings as used in the shrinkTVP R package \cite{Knaus2021}.
\begin{eqnarray}
\beta_{p,m,k} \mid \beta_{p,m,k-1}, \sigma_{\beta_{m}} \sim N(\beta_{p,m,k-1}, \sigma_{\beta_{m}}^{2}) \\
\sigma_{\beta_{m}} \mid \xi_{m}^{2} \sim N^{+}(0, \xi_{m}^{2}) \\
\xi_{m}^{2} \mid a^{\xi}, \kappa_{m}^{2} \sim Gamma(a^{\xi}, \frac{a^{\xi}\kappa_{m}^{2}}{2}) \\
\kappa_{m}^{2} \mid c^{\xi}, \kappa_{B}^{2} \sim Gamma(c^{\xi}, \frac{c^{\xi}}{\kappa_{B}^{2}}) \\
2a^{\xi} \sim Beta(5,10) \\ 
2c^{\xi} \sim Beta(5,2) \\ 
\frac{\kappa_{B}^{2}}{2} \sim F(1,1)
\label{eq:S8eq}
\end{eqnarray}
In addition this B-spline model, we also considered a multivariate normal B-spline. We begin using the same formation for $\psi_{p,t,m}$ as given in equation \ref{eq:S5eq}. In this model, we estimate the spline coefficients and intercept parameters using multivariate Normal priors. The estimation of the intercept $\alpha_{p, m}^{prov.}$ uses the same hierarchical multivariate normal priors as described in equations \ref{eq:eq5} to \ref{eq:eq6} of the main paper. For the estimation of the spline coefficients, we use a multivariate Normal prior such that, 
\begin{eqnarray}
\beta_{p,1:M,k} \mid \beta_{p,1:M,k-1}, \Sigma_{\beta} \sim MVN(\beta_{p,1:M,k-1}, \Sigma_{\beta}).
\label{eq:S9eq}
\end{eqnarray}
As before, we use a sum-to-zero constraint on the spline coefficients to ensure idenitifability and a vague Wishart prior to estimate the variance-covariance matrix of the spline coefficients:
\begin{eqnarray}
\Sigma_{\beta} \sim Wishart(I_{M}, M+1).
\label{eq:S10eq}
\end{eqnarray}


\nolinenumbers

%
%
%
\clearpage

\bibliography{plos_latex_template}

\end{document}